# Observation of two-dimensional superconductivity at the LaAlO$_3$/KTaO$_3$(110) heterointerface

Zheng Chen[1], Zhongran Liu[2], Yanqiu Sun[1], Xiaoxin Chen[2], Yuan Liu[1], Hui Zhang[3], Hekang Li[1], Meng Zhang[1], Siyuan Hong[1], Tianshuang Ren[1], Chao Zhang[4], He Tian[2], Yi Zhou[3,5,6], Jirong Sun[3,5], and Yanwu Xie[1,7]*

[1]Interdisciplinary Center for Quantum Information, Zhejiang Province Key Laboratory of Quantum Technology and Device, Department of Physics, Zhejiang University, Hangzhou 310027, China
[2]Center of Electron Microscope, State Key Laboratory of Silicon Materials, School of Materials Science and Engineering, Zhejiang University, Hangzhou, 310027, China
[3]Beijing National Laboratory for Condensed Matter Physics & Institute of Physics, Chinese Academy of Sciences, Beijing 100190, China
[4]Instrumentation and Service Center for Physical Sciences, Westlake University, Hangzhou 310024, China
[5]Songshan Lake Materials Laboratory, Dongguan, Guangdong 523808, China
[6]Kavli Institute for Theoretical Sciences and CAS Center for Excellence in Topological Quantum Computation, University of Chinese Academy of Sciences, Beijing 100190, China
[7]Collaborative Innovation Center of Advanced Microstructures, Nanjing University, Nanjing 210093, China

*To whom correspondence should be addressed. E-mail: ywxie@zju.edu.cn

**Abstract.**

We report on the observation of a $T_c$ ~0.9 K superconductivity at the interface between LaAlO$_3$ film and the 5$d$ transition-metal oxide KTaO$_3$(110) single crystal. The interface shows a large anisotropy of the upper critical field, and its superconducting transition is consistent with a Berezinskii-Kosterlitz-Thouless transition. Both facts suggest that the superconductivity is two-dimensional (2D) in nature. The carrier density measured at 5 K is ~7 × 10$^{13}$ cm$^{-2}$. The superconducting layer thickness and coherence length are estimated to be ~8 and ~30 nm, respectively. Our result provides a new platform for the study of 2D superconductivity at oxide interfaces.

Oxide interfaces exhibit a rich variety of emergent phenomena [1]. One remarkable observation is the two-dimensional (2D) superconductivity that exists at interfaces between two non-superconducting constituents [2–9]. Two well-known examples are the LaAlO$_3$/SrTiO$_3$(LAO/STO) heterostructure [2] and the La$_2$CuO$_4$/La$_{2-x}$Sr$_x$CuO$_4$ bilayer [4]. In both cases, the formation of interface induces charge carriers (electrons or holes) in the parent compound (STO or La$_2$CuO$_4$) and confines them near the interface. These superconducting interfaces provide an ideal ground for studying 2D superconductivity, and are relevant to understanding the high-temperature superconducting copper oxides and for the development of superconductor-based devices.

In previous studies [2,3,6,9–12], the STO-based superconducting interfaces, particularly LAO/STO, have attracted considerable interest, although their critical temperature, $T_c$, is low (< 0.3 K). The parent compound STO is a wide-band insulator (semiconductor) and is the first known superconducting semiconductor [13]. One important feature of STO-based interface superconductivity is its high controllability by applying a gate voltage [3,14,15], owing to its low carrier density and the large dielectric constant of STO. In addition, it exhibits a few remarkable properties such as the coexistence with ferromagnetism [15–17], the high-temperature-superconductor-like gap behavior [18], and the multiple quantum criticality [10]. In addition to STO, KTaO$_3$ (KTO) is another versatile oxide that has attracted much attention due to its interesting dielectric, photoconductive, and optical properties [19]. KTO is in many ways similar to STO [20,21]. Both of them are of perovskite structure, are incipient ferroelectrics characterized by extremely large dielectric constants at low temperatures, and have similar band structures [22]. While the conduction band of STO is in a 3$d$ band associated with the titanium ions, the conduction band of KTO is in a 5$d$ band associated with the tantalum ions. In analogy with the STO-based interfaces, the KTO-based interfaces can also host 2D electron gases [23–25]. However, unlike STO, superconductivity has never been reported in chemically doped KTO [20,26]. It was only reported in the surface of KTO(001) single crystal gated by an electric double-layer technique, with an extremely low $T_c$ ~50 mK [26]. Surprisingly, very recently Liu *et al.* [27] observed superconductivity with a high $T_c$ (~2 K) at the interfaces between EuO (or LAO) and (111)-oriented KTO single crystal. By contrast, they detected no superconductivity (down to 25 mK) at the corresponding KTO(001) interfaces [27]. In addition to (001) and (111), (110) is another principal orientation for perovskite structure. In this work, we report on the existence of 2D superconductivity ($T_c$ ~0.9 K) at (110)-oriented KTO interfaces.

Our samples were prepared by depositing LAO films on KTO(110) single crystal substrates by pulsed laser deposition[28]. The typical growth temperature is 620 °C; the typical growth atmosphere is 1 × 10$^{-5}$ mbar O$_2$ (adding a tiny amount of H$_2$O vapor, 1 × 10$^{-7}$ mbar). After growth, the samples were cooled to room temperature under the same atmosphere. Atomic force microscopy characterizations (Fig. S1) show that the surfaces of both KTO substrates and LAO films are very smooth, with a root-mean square roughness of ~0.2 nm. X-ray diffraction shows no epitaxial peaks of LAO (Fig.

S2), indicating that the LAO film is not in a well-crystalline state. Scanning transmission electron microscopy (STEM) high-angle annular dark-field (HAADF) images taken from a 20-nm LAO/KTO(110) heterostructure are given in Figs. 1a & S3a. The interface is abrupt and smooth. The KTO substrate near the interface is in a well-crystalline state, while the LAO film is amorphous. Note that the LAO films grown on KTO(001) [23] and KTO(111)[27] substrates were also found to be amorphous, probably due to the large lattice mismatch between LAO (0.379 nm) and KTO (0.399 nm). Atomic-scale energy-dispersive X-ray spectroscopy (EDS) elemental mapping was performed on the same sample (Figs. 1b & S3b). The EDS maps show that the top-most K-layer is missing, forming one-unit-cell La-Ta layer. Except for this layer, the interface is chemically abrupt, and the atomic intermixing is limited to about one unit cell. X-ray photoemission spectroscopy (XPS) core-level measurements (Fig. S4) confirm that there exists no metallic Al or Ta in our samples.

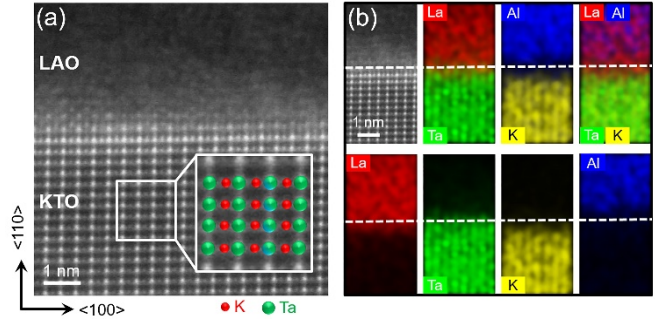

FIG. 1. Structural characterization of a 20-nm LAO/KTO(110) heterostructure. (a) HAADF-STEM image shows that the LAO film is amorphous. The inset shows the image with a higher magnification and the atomic configuration (colored) of KTO. (b) EDS elemental mapping of the same sample. The top-left corner panel is the HAADF image of the region for EDS mapping. For comparison, the EDS mapping images of different elements (La, Ta, Al, and K) are presented both together and separately. The interface is abrupt, and intermixing is not significant. At the interface, about one monolayer of K is missing, and as a result, a La-rich layer is formed. The dashed line indicates the position of interface.

We examined the electrical transport properties of the LAO/KTO(110) samples. More than 10 samples have been measured, ensuring reproducibility. In the following, we present the results of two typical samples that have been pre-patterned into a 20-μm-wide Hall-bar configuration (Figs. 2a & 2b) [28]. The samples are metallically conducting in a wide temperature range (Fig. 2c), indicating the formation of electron gas at the interface. The magnetic-field-dependent Hall resistance, $R_{\text{Hall}}$, measured at $T = 5$ K (Fig. 2d) confirms that the charge carriers are electrons; the 2D Hall carrier densities, $n_{\text{2DS}}$, of the 6-nm and 20-nm samples are $7.1 \times 10^{13}$ cm$^{-2}$ and $6.7 \times 10^{13}$ cm$^{-2}$, and the mobilities 81.6 cm$^2$V$^{-1}$s$^{-1}$ and 80.8 cm$^2$V$^{-1}$s$^{-1}$, respectively. It is evident that the electron gas locates in KTO, rather than LAO, near the interface, because the bandgap of LAO (5.6 eV) is much larger than that of KTO (3.6 eV). This conclusion is supported by many previous studies which have already showed that the LAO films are highly insulating [29–31], while electron gas can be formed on the surface/interface of KTO [23,25,26,32,33]. Interestingly, the normal-state sheet resistance, $R_{\text{sheet}}(T)$, shows a nearly linear dependence on $T$, which is apparently similar to that observed for the La$_2$CuO$_4$/La$_{2-x}$Sr$_x$CuO$_4$ bilayer [4], but dissimilar to the LAO/STO heterostructure [2] and the LAO/KTO(001) [23] or LaTiO$_3$/KTO(001) [25] heterostructures. This kind of $R_{sheet}(T)$ characteristic is indicative of a non-Fermi liquid behavior [34]. Further study is needed to verify it, which is beyond the scope of the present work.

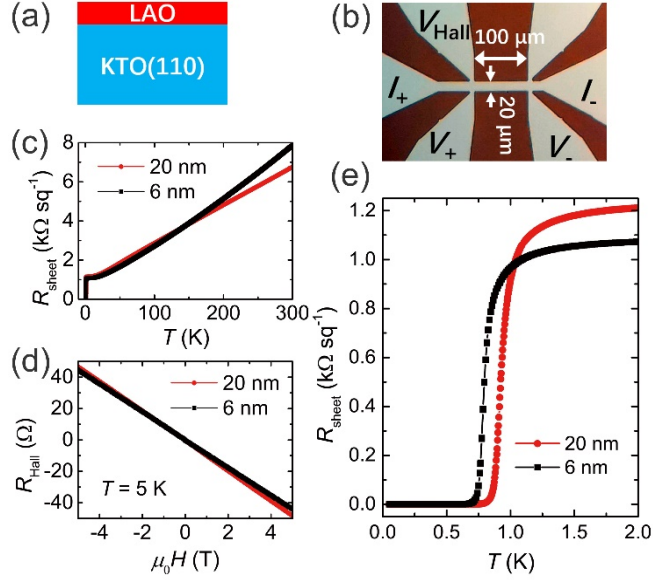

FIG. 2. (a) Schematic view of the samples. (b) A photo of the central Hall bar area. The interface of the Hall-bar region (light) is conducting; other regions (dark) are insulating. (c) Dependence of $R_{sheet}$ on $T$ of the 6-nm and 20-nm samples measured in a wide temperature range. (d) Dependence of $R_{Hall}$ on magnetic field $\mu_0 H$. (e) Dependence of $R_{sheet}$ on $T$ in low temperatures.

As the origin of the electron gas is concerned, in analogy with STO-based interfaces [30,31,35–38], we consider three possibilities: polar-discontinuity-induced electronic reconstruction [36], interface chemistry (*i.e.*, the substitution of K with La) [37], and oxygen vacancies (in both KTO and LAO) [30,31,35]. The polar-discontinuity and other structure-related scenarios can be ruled out because the LAO film is amorphous. The interface chemistry can also be fairly ruled out because STEM and EDS characterizations demonstrate that the interface intermixing is not significant (Figs. 1 & S3). Thus, the most likely scenario is the oxygen vacancies. This scenario is supported by the fact that the LAO/KTO(110) interfaces are less conducting if they have been grown or post-annealed in an oxygen-rich environment (Fig. S5). In addition, we found that adding a tiny amount of water vapor can enhance the interface conductance (Fig. S6), presumably due to the interaction of water vapor with oxygen vacancies in LAO [30].

Next, we turn to the occurrence of interface superconductivity. As shown in Figs. 2c &2e, at low temperatures the $R_{sheet}(T)$ curves of both samples drop sharply, undergoing a transition to the superconducting state. The zero-resistance state is observed at ~0.54 and ~0.59 K, respectively, for the 6-nm and 20-nm samples. The mid-point $T_c$s, defined as 50% normal state resistance, are 0.80 and 0.93 K for the 6-nm and 20-nm samples, respectively. The widths of the transitions (20% to 80%) of the 6-nm and 20-nm samples are 0.11 and 0.12 K, respectively. The mid-point $T_c$ observed here is nearly 20 times higher than that in the electric double-layer-gated KTO(001) single crystal [26], and around 3 times higher than that in LAO/STO interfaces [2,11,12].

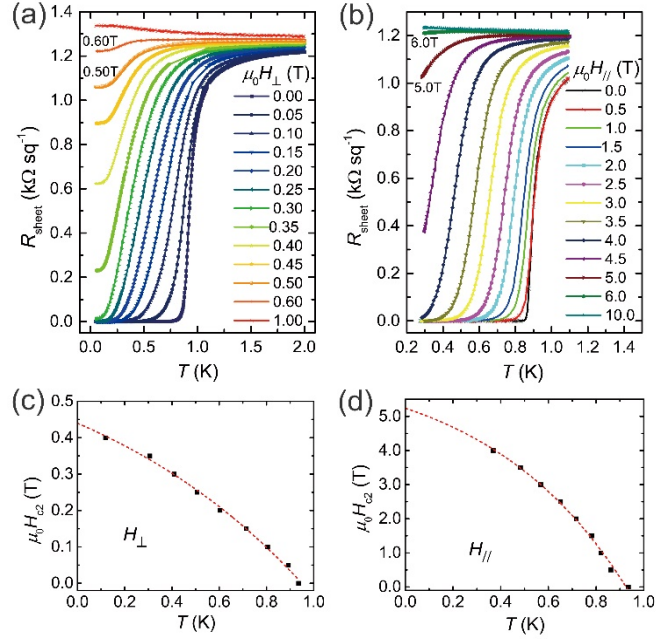

FIG. 3. Transport behaviors under magnetic field for the 20-nm sample. Dependence of $R_{sheet}$ on $T$ for field (a) perpendicular and (b) parallel to the interface. Temperature dependence of upper critical field $\mu_0 H_{c2}$, extracted from the 50% normal state resistance, for (c) perpendicular and (d) parallel to the interface. The extrapolated $\mu_0 H_{c2}(0)$s agree fairly well with that measured in the low-temperature magnetic-field-dependent resistance (Fig. S7).

To investigate the nature of this interface superconductivity further, we measured the temperature-dependent $R_{sheet}(T)$s under magnetic fields applied perpendicular and parallel to the interface. The result for the 20-nm sample is shown in Figs. 3a & 3b. We found that application of a magnetic field, $\mu_0 H$ (here $\mu_0$ is the vacuum permeability), of ~0.6 T perpendicular (or ~6.0 T parallel) to the interface is needed to suppress the superconducting state. This strong anisotropy suggests that the superconductivity is 2D. Figures 3c & 3d show the temperature-dependent upper critical field, $\mu_0 H_{c2}$, derived from the $R_{sheet}(T)$ curves in Figs. 3a & 3b. The Ginzburg-Landau coherence length, $\xi_{GL}$, can be extracted using the linearized Ginzburg-Landau form [39,40] $\mu_0 H_{c2}^{\perp}(T) = \frac{\phi_0}{2\pi \xi_{GL}^2(0)}(1 - \frac{T}{T_c})$, where $\phi_0$ is the flux quantum and $\xi_{GL}(0)$ is the extrapolation of $\xi_{GL}$ to $T = 0$. Using the extrapolated $\mu_0 H_{c2}^{\perp}(0) = 0.44$ T (Fig. 3c), we obtained $\xi_{GL}(0) = 27.3$ nm. For a 2D superconductor, $\mu_0 H_{c2}^{\parallel}(T) = \frac{\phi_0 \sqrt{12}}{2\pi \xi_{GL}(0) d_{SC}}(1 - \frac{T}{T_c})^{1/2}$, where $d_{sc}$ is the superconducting thickness [39,40]. Using the extrapolated $\mu_0 H_{c2}^{\parallel}(0) = 5.24$ T (Fig. 3d), we obtained $d_{sc} = 8.0$ nm. With similar analysis, we obtained $\xi_{GL}(0)$ and $d_{sc}$ to be 41.0 and 7.7 nm, respectively, for the 6-nm sample. In both samples, $d_{sc}$ is much smaller than $\xi_{GL}(0)$, supporting that the superconductivity is 2D in nature.

Assuming a single-band model, the mean free path $l_{mfp}$ of the conducting electrons can be estimated using $k_F = \sqrt{2\pi n_{2D}}$ and $l_{mfp} = \frac{h}{e^2}\frac{1}{k_F R_{sheet}}$, where $k_F$ and $h$ are Fermi wave number and Planck's constant, respectively [40]. From the measured $n_{2D}$(5 K) and $R_{sheet}$(5 K), the $l_{mfp}$ of the present interfaces is estimated to be ~10 nm. This value is comparable with the superconducting layer thickness $d_{sc}$, and is about 1/4 to 1/3 of $\xi_{GL}$, suggesting that the superconductivity is in an intermediate range between clean ($l_{mfp} > \xi_{GL}$) and dirty ($\xi_{GL} \gg l_{mfp}$) limits.

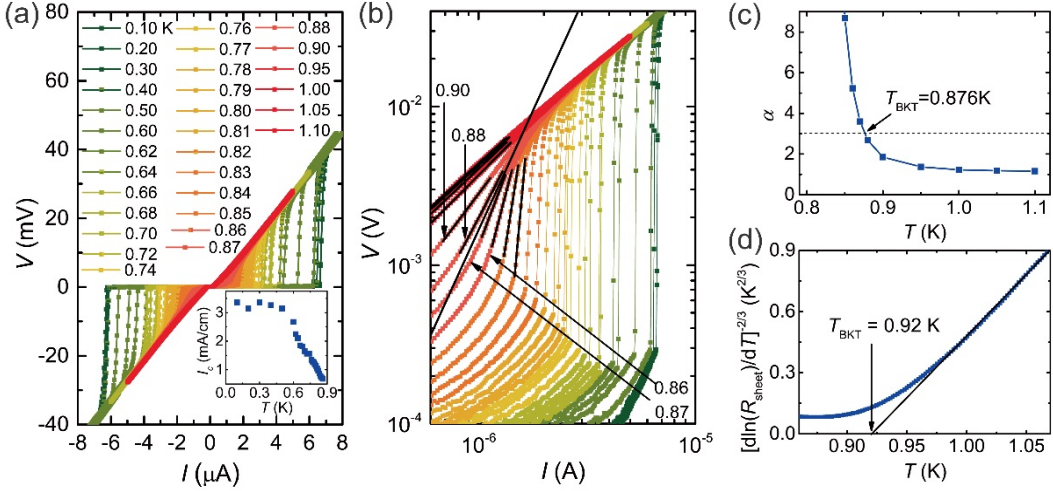

FIG. 4. (a) Temperature-dependent $I$-$V$ measurements for the 20-nm sample (the measured bridge is 20 × 100 μm², the central region of the Hall bar shown in Fig. 2b). Inset: Temperature dependence of the linear critical current density, $I_c$. (b) $I$-$V$ curves on a logarithmic scale. The symbol label is the same as that in (a). The long black line corresponds to $V \sim I^3$ dependence and shows that 0.87 K < $T_{BKT}$ < 0.88 K. (c) Temperature dependence of the power-law exponent α, as deduced from the fits shown in (b). (d) $R_{sheet}(T)$ dependence of the same sample (the same data shown in Fig. 2d), plotted on a $[d\ln(R_{sheet})/dT]^{-2/3}$ scale. The solid line is the behavior expected for a BKT transition with $T_{BKT}$ = 0.92 K.

To gain further insights of the superconductivity at the LAO/KTO(110) interface, we carried out current-voltage ($I$-$V$) measurements. The data for the 20-nm sample are shown in Fig. 4. Below $T_c$ we find a clear critical current, $I_c$, whose value decreases with increasing measurement temperature (Fig. 4a and its inset). This is one more evidence for the existence of superconductivity at the interface. The maximal value of $I_c$ is ~3.4 mA/cm, which is more than 30 times higher than that observed for the superconducting LAO/STO interface [2].

For a 2D superconductor, the transition into the superconducting state is a Berezinskii-Kosterlitz-Thouless (BKT) transition, characterized by a transition temperature $T_{BKT}$ that corresponds to the unbinding of vortex-antivortex pairs [2,41]. The BKT behavior would result in a $V \propto I^\alpha$ power-law dependence, with $\alpha(T_{BKT}) = 3$, in *I-V* curves. As shown in Fig. 4b, the power-law $V \propto I^\alpha$ dependence is seen. At $T = 0.876$ K, the exponent α approaches 3 (Fig. 4c); this temperature is thus defined as $T_{BKT}$. In addition, close to $T_{BKT}$, a $R_{sheet}(T) = R_0\exp[-b(T/T_{BKT}-1)^{-1/2}]$ dependence, where $R_0$ and $b$ are material parameters, is expected [42]. As shown in Fig. 4d, the measured $R_{sheet}(T)$ is consistent with this dependence and yields $T_{BKT} = 0.92$ K, in agreement with the analysis of α exponent. Therefore, the superconducting transition of the LAO/KTO(110) samples is consistent with that of a 2D superconducting film.

Similar to LAO/STO interfaces [11,12,14], Pauli paramagnetic limit is violated in the present LAO/KTO(110) interfaces. For a weak coupling BCS superconductor, Pauli paramagnetic limit sets the upper bound for the parallel critical field [12,43,44], which is given by $\mu_0 H_{c2}^P \approx 1.76 k_B T_c/\sqrt{2}\mu_B$, where $k_B$ and $\mu_B$ are the Boltzmann's constant and Bohr magneton, respectively. Using $T_{BKT} = 0.92$ K as the superconducting transition temperature $T_c$, we obtained a $\mu_0 H_{c2}^P$ of 1.70 T, which is only about 1/3 of the measured $\mu_0 H_{c2}^\parallel$ (~5.24 T). In LAO/STO interfaces the violation was proposed to be a consequence of strong spin-orbit coupling [12,14]. A similar mechanism is expected for the LAO/KTO(110) interfaces since KTO has a strong spin-orbit coupling strength due to the 5*d* tantalum atoms [24,32,45].

Finally, we discuss the possible origin of the interface superconductivity at LAO/KTO(110) interfaces. The present experiments do not allow us to determine the underneath mechanism yet. However, a few useful points may be given. (1) Parasite superconductivity from Al or Ta metal can be ruled out by structural characterizations (Figs. 1, S3, and S4 ) and also by the observation of relatively large upper critical fields (Figs. 3 and S7. Note that the upper critical fields of Ta and Al metals are well below 0.1 T). (2) The determined thickness of superconducting layer is ~8 nm, which rules out the possibility that the superconductivity originates from the interface La-Ta layer (~0.5 nm, Figs. 1 and S3). (3) Because LAO is amorphous, strain effect should be of less relevance [46]. (4) Because the measured Hall mobility is more than one order lower than that of electron-doped single-crystal KTO surfaces [26,33], the superconducting layer should contain relatively high-density disorders, most likely from oxygen vacancies, given that the interface intermixing is not significant.

Thus, the above points lead to a conclusion that the observed superconductivity is intrinsic to the interface KTO layer that is of a few nanometers. Since the bulk KTO is not known to be superconducting, we speculate that here the surface (interface) bands or phonons [47,48], as suggested by Ginzburg [47], play a determinant role. Due to various factors such as symmetry breaking, atomic/orbital reconstruction, charge transfer, and quantum confinement, both the energy bands and phonon spectra of the interface KTO could be very different from that of the bulk KTO, which may result in metallic character and attraction between carriers [47,48]. Following this speculation, the strong dependence of superconductivity on the KTO orientation, as revealed by Liu's [27] and the present observations (also see Fig. S8 for LAO/KTO interfaces with three different orientations), can be explained by the possible orientation-dependent surface bands and phonon spectra. In addition, a simple comparison (Table S1) of the superconductivity at LAO/STO, EuO/KTO(111), and LAO/KTO(110) interfaces implies that higher $T_c$ is favored by higher $n_{2D}$, thinner superconducting layer thickness, and higher density of disorders, which seems to partially support that electron-phonon coupling has a strong influence on the interface superconductivity. Further studies are needed for deeper understanding.

In conclusion, we have demonstrated that a conducting electron gas can be formed at the interface between LAO film and KTO(110) single crystal, which becomes superconducting at low temperatures with a $T_c$ up to 0.9 K. This interface superconductivity is found to be 2D in nature, by the large anisotropy of the upper critical magnetic field and by the consistence with the BKT transition behaviors. The superconducting layer thickness and the coherence length are estimated to be ~8 nm and ~30 nm, respectively. Our present discovery provides a new model system for studying the rich physics of 2D superconductivity and other emergent phenomena at oxide interfaces.

**Acknowledgments.** We thank H. Y. Hwang for valuable discussion. This work was supported by the National Key Research and Development Program of China (2017YFA0303002, 2016YFA0300204, 2018YFA0305704), National Natural Science Foundation of China (11934016, 11520101002), Zhejiang Province Plan for Science and Technology (2020C01019), and the Fundamental Research Funds for the Central Universities of China. J.S. thanks the support of the Project for Innovative Research Team of National Natural Science Foundation of China (11921004).

# Supplemental Material for

# Observation of two-dimensional superconductivity at the LaAlO$_3$/KTaO$_3$(110) heterointerface

Zheng Chen, Zhongran Liu, Yanqiu Sun, Xiaoxin Chen, Yuan Liu, Hui Zhang, Hekang Li, Meng Zhang, Siyuan Hong, Tianshuang Ren, Chao Zhang, He Tian, Yi Zhou, Jirong Sun, and Yanwu Xie*

*Correspondence to: ywxie@zju.edu.cn (YX)

**This file includes:**

    Materials and Methods
    Figs. S1 to S8
    Table S1
    Reference

**Materials and Methods.**

**Samples.** The samples were prepared by depositing LAO films on KTO(110) single crystalline substrates (MTI Corporation) in a pulsed laser deposition chamber. A 248-nm KrF excimer laser was used. The laser fluence is ~1 Jcm$^{-2}$; the laser repetition rate is 2 Hz. The LAO target is a single crystal. The depositions were performed at 620 °C, in a mixed atmosphere of $1\times10^{-5}$ mbar $O_2$ and $1\times10^{-7}$ mbar $H_2O$ vapor. The addition of a tiny amount of $H_2O$ vapor is found to be helpful for the formation of the conducting electron gas at the interface (Fig. S6). The film thickness was controlled by counting laser pulses and the growth rate was calibrated to be ~0.02 nm/laser pulse. After growth, the samples were cooled down to room temperature under the growth atmosphere.

**Hall bar.** Hall bars were pre-patterned onto KTO substrates by a hard-mask method [1] using standard optical lithography and lift off techniques. As shown in Fig. 2b, the hard mask (dark regions) was made of ~100-nm-thick AlO$_x$ films, grown by pulsed laser deposition at room temperature, in $P(O_2) = 0.01$ mbar. After patterning, the active Hall-bar area is the uncovered KTO(110) surface. These patterned substrates were used for growth as described above. After growth of LAO, only the active Hall-bar area (light region in Fig. 2b) is conducting. By contrast, the area covered by AlO$_x$ hard mask is highly insulating.

**Scanning transmission electron microscopy (STEM) and Energy-dispersive X-ray spectroscopy (EDS) mapping.** Cross-sectional specimens for electron microscopy investigations were prepared by a FEI quanta 3D FEG Focused Ion Beam. Atomic resolution HAADF-STEM images and EDS mappings were acquired by a spherical aberration corrected electron microscope equipped with 4 Super-X EDS detectors (FEI Titan G2 80-200 ChemiSTEM).

**X-ray photoemission spectroscopy (XPS) measurements.** The XPS measurements were performed at room temperature with the laboratory monochromatic Al Kα X-ray source. The samples were transferred into the XPS chamber through air, and the measurements were performed without any in-situ cleaning.

**Transport measurements.** The contacts to the conducting LAO/KTO interfaces were made by ultrasonic bonding with Al wires. The transport measurements were carried out in a commercial physical property measurement system (PPMS, Quantum Design) with a dilution refrigerator insert and in a commercial $^4$He cryostat with a $^3$He insert (Cryogenic Ltd.).

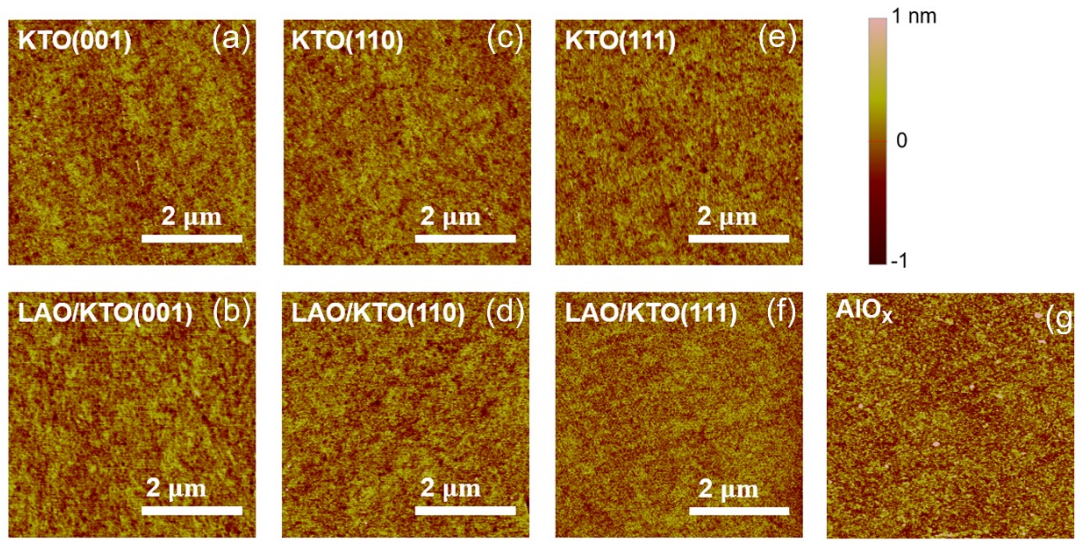

FIG. S1. Surface morphologies of KTO substrates, LAO/KTO samples, and AlO$_x$ hard mask characterized by atomic force microscopy. All the surfaces are very smooth. The measured root mean square roughness (over 5 × 5 μm$^2$) from (a) to (g) is 0.21, 0.19, 0.18, 0.20, 0.18, 0.19 and 0.34 nm, respectively. The LAO thicknesses are 6 nm in (b, d) and 20 nm in (f). The AlO$_x$ thickness is ~100 nm in (g).

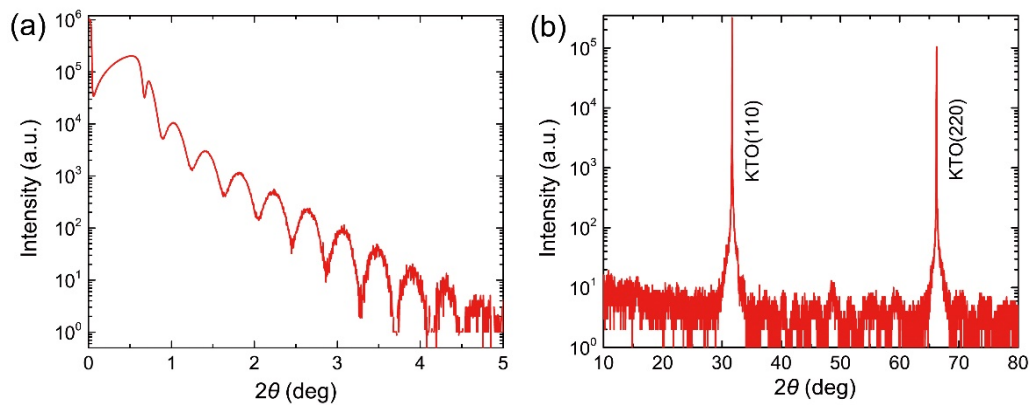

FIG. S2. XRD characterization of 20-nm LAO/KTO(110) sample. (a) Small-angle X-ray reflectivity measurement. Fitting of the oscillations in the curve give a thickness of ~20 nm. (b) The $\theta$-$2\theta$ scan of the sample. Both peaks are from the KTO substrate. No epitaxial peaks from LAO film were detected.

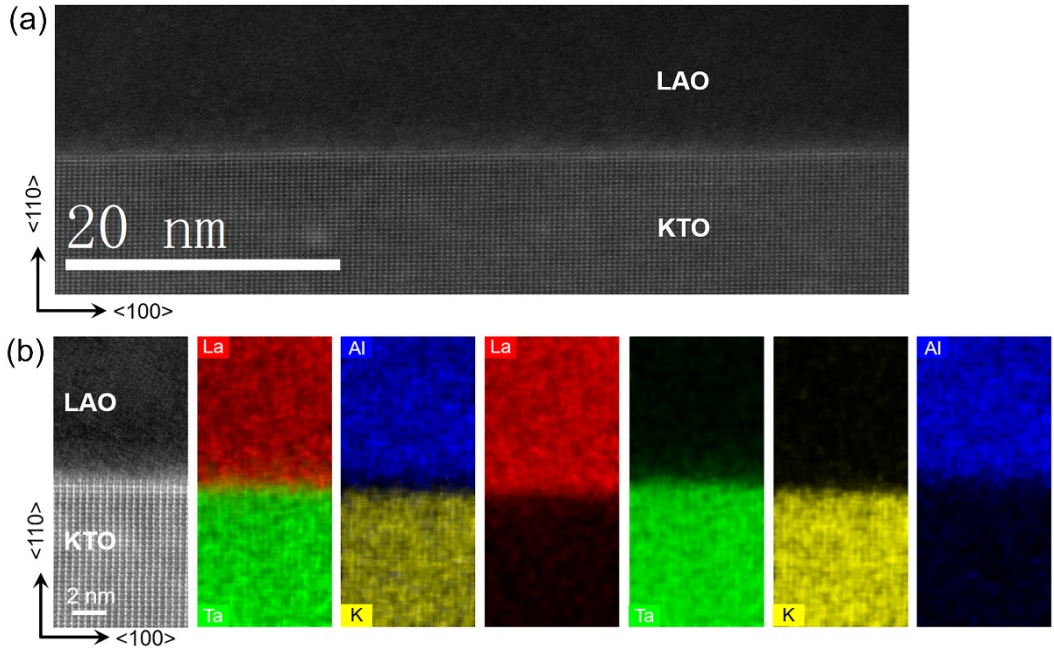

FIG. S3. Structural characterization of the same 20-nm LAO/KTO(110) heterostructure in a larger scale. (a) HAADF-STEM image shows that the LAO film is amorphous and homogeneous, and the interface is abrupt. (b) EDS elemental mapping shows that the interface is abrupt, and intermixing is not significant. The left-most panel is the HAADF image of the region for EDS mapping. For comparison, the EDS mapping images of different elements (La, Ta, Al, and K) are presented both together and separately.

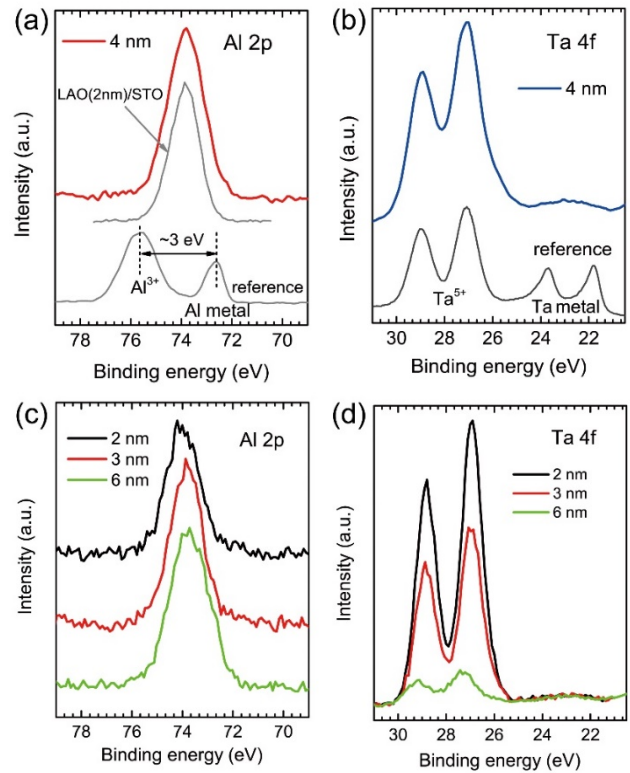

FIG. S4. XPS core-level spectra of LAO/KTO(110) samples. (a) Al 2p spectra (red line) of a 4-nm sample. The reference spectra is from Al metal with $Al_2O_3$, which shows that the binding energy of Al metal is about 3 eV lower than that of Al$^{3+}$. The single-peak feature observed here suggests that all the Al elements are in oxidation state. The apparent shift of the Al 2p peak compared with the reference spectra is attributed to their different local chemical environments. To verify this point, we also measured the Al 2p spectra of a 2-nm epitaxially grown LAO/STO heterostructure. As shown in the figure, the Al 2p spectra of LAO/STO and LAO/KTO coincide excellently. (b) Ta 4f spectra (blue line) of the same 4-nm sample. The reference spectra is from Ta metal with Ta$_2$O$_5$. The broad bump around 23 eV is due to O 2s core level. (c) Al 2p and (d) Ta 4f core-level spectra of LAO/KTO(110) samples with other LAO thicknesses. The XPS data presented here show that there are no Al or Ta metals existing in our samples.

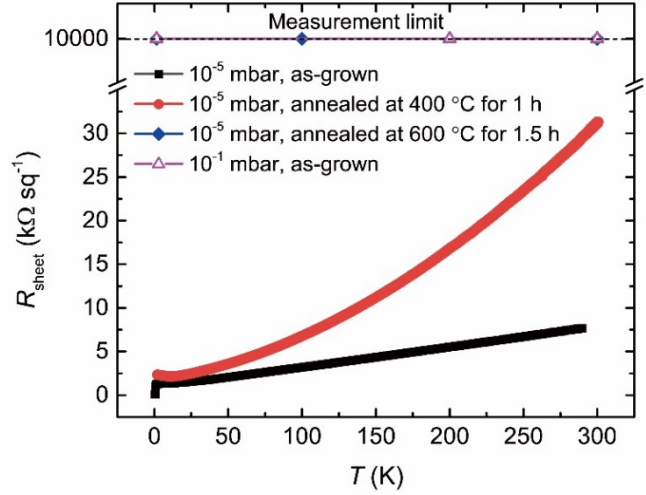

FIG. S5. Temperature-dependent $R_{sheet}$ for 20-nm LAO/KTO(110) samples. The oxygen pressure during growth (before comma) and the post-annealing condition (after comma) are as labelled in the figure. Other growth conditions are the same as that described in the "Samples" section. A high oxygen pressure of 200 mbar was used for post annealing. These data suggest that oxygen vacancies play a dominant role on the interface conductance.

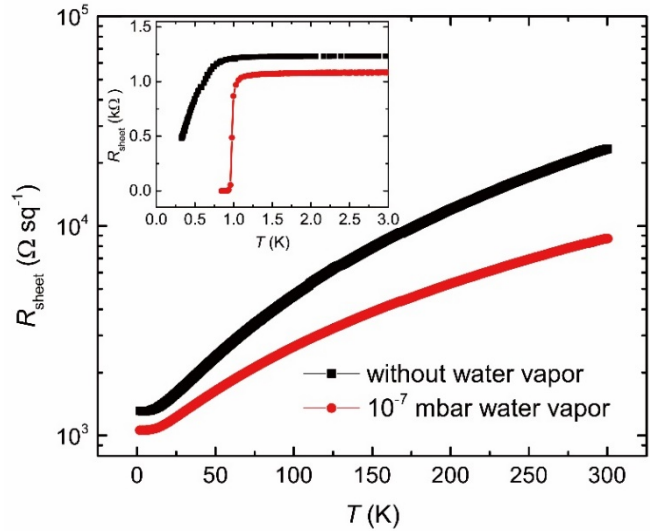

FIG. S6. Temperature-dependent $R_{sheet}$ for 6-nm LAO/KTO(110) samples grown with (red circles) and without water vapor (black squares), in a temperature range from 2 to 300 K. Inset: The $R_{sheet}(T)$ data for $T \leq 3$ K.

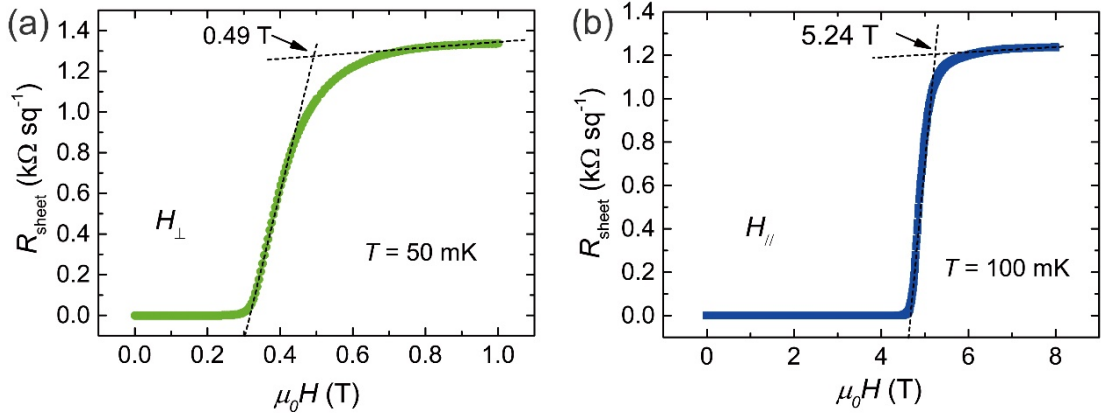

FIG. S7. Magnetic field dependent $R_{sheet}$ of the 20-nm LAO/KTO(110) sample measured at an extreme low temperature for a field (a) perpendicular and (b) parallel to the interface.

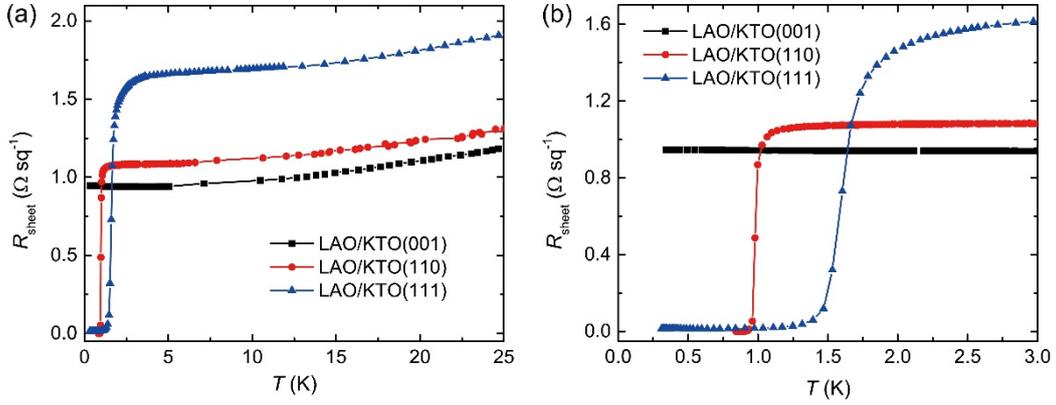

FIG. S8. Comparison of temperature-dependent $R_{sheet}$ for LAO/KTO(001), LAO/KTO(110), and LAO/KTO(111) heterointerfaces prepared under similar conditions.

Table S1. A few typical parameters for the superconductivity at different interfaces.

|  | $n_{2d}$ ($\times 10^{13}$ cm$^{-2}$) | $\mu_{Hall}$ (cm$^2$V$^{-1}$s$^{-1}$) | $T_c$ (K) | Superconducting layer thickness (nm) | Coherence length (nm) |
| --- | --- | --- | --- | --- | --- |
| LAO/STO [2] | ~4 | ~350 | ~0.3 | ~10 | ~70 |
| EuO/KTO(111) [3] | ~10 | -- | ~2.0 | ~5 | ~13 |
| LAO/KTO(110) (the present one) | ~7 | ~81 | ~0.9 | ~8 | ~30 |

**Reference.**

[1] Y. Xie, Y. Hikita, C. Bell, and H. Y. Hwang, Nat. Commun. **2**, 494 (2011).

[2] N. Reyren, S. Thiel, A. D. Caviglia, L. Fitting Kourkoutis, G. Hammerl, C. Richter, C. W. Schneider, T. Kopp, A. S. Rüetschi, D. Jaccard, M. Gabay, D. A. Muller, J. M. Triscone, and J. Mannhart, Science **317**, 1196 (2007).

[3] C. Liu, X. Yan, D. Jin, Y. Ma, H.-W. Hsiao, Y. Lin, T. M. Bretz-Sullivan, X. Zhou, J. Pearson, B. Fisher, J. S. Jiang, W. Han, J.-M. Zuo, J. Wen, D. D. Fong, J. Sun, H. Zhou, and A. Bhattacharya, arXiv: 2004.07416 (2020).